\documentclass{iopconfser}
\usepackage{graphicx}
\usepackage{subfig}
\usepackage{amsmath}
\usepackage{float}
\usepackage{lineno}

\begin{document}

\title{LHC beam monitoring via real-time hit reconstruction in the LHCb VELO pixel detector }

\author{Daniele Passaro$^{1,2}$, Giulio Cordova$^{2,3}$, Federico Lazzari$^{2,3}$, Elena Graverini$^{2,3,4}$, Micheal Morello$^{1,2}$ and Giovanni Punzi$^{2,3}$}

\affil{$^1$Scuola Normale Superiore, Pisa, Italy}
\affil{$^2$INFN Sezione di Pisa, Pisa, Italy}
\affil{$^3$Dipartimento di Fisica ``Enrico Fermi", Università degli Studi di Pisa, Pisa, Italy}
\affil{$^4$Institute of Physics, Ecole Polytechnique F\'ed\'erale de Lausanne, Lausanne, Switzerland}

\email{daniele.passaro@sns.it}

\begin{abstract}
The increasing computing power and bandwidth of programmable digital devices opens new possibilities in the field of real-time processing of HEP data. The LHCb collaboration is exploiting these technology advancements in various ways to enhance its capability for complex data reconstruction in real time. Amongst them is the real-time reconstruction of hits in the VELO pixel detector, by means of real-time cluster-finding embedded in the readout board firmware. This reconstruction, in addition to saving data-acquisition bandwidth and high-level trigger computing resources, also enables further useful applications in precision monitoring and diagnostics of LHC beam conditions. In fact, clusters of pixels, being more reliable and robust indications of physical particle hits than raw pixel counts, are also exempt from the complications associated to the reconstruction of tracks, that involves alignment issues and is sensitive to multi-layer efficiency products. In this paper, we describe the design and implementation of a flexible system embedded in the readout firmware of the VELO detector, allowing real-time measurement of cluster density in several parts of the detector simultaneously, and separately for every bunch ID, for every single LHC collision, without any slowdown of data acquisition. Quantitative applications of this system to luminosity measurement and beam monitoring are demonstrated.
\end{abstract}

\vspace{-0.5cm}
\section{Leveraging the FPGA power}
The usefulness to reconstruct relevant physical proxies, such as particle hits and tracks, at the earliest stages of the DAQ in order to speed up the reconstruction stages, is now becoming widely recognised \cite{het-compu}.
The LHCb-Upgrade I experiment \cite{lhcbcollaboration2023lhcb} has adopted for the first time a trigger system based on the complete reconstruction of all collision events. The real-time reconstruction of complex events at an average rate of 30 MHz is a very significant computational challenge, which the LHCb experiment has decided to tackle with a heterogeneous computing system, in which CPU, GPU and Field Programmable Gate Arrays (FPGA) cards are employed at the same time. The FPGAs are unique in this sense, as they can process a huge amount of data at the readout level and in a fully parallel way, relieving the High-Levels Triggers of logically simple but time-consuming tasks.
Many R\&D projects have been established in LHCb to explore the usage of heterogeneous computing systems to enhance the computing power of the experiment. Amongst them, we discuss the Retina project \cite{Lazzari:2888549, LHCbcollaboration:2886764} here. Its main goal is to implement on FPGAs, at the readout level of the DAQ, a parallel, high throughput, low latency tracking algorithm, named \textit{Artificial Retina}. The first stage of the Retina algorithm is the reconstruction of hits on the detector sensors.
The availability of high-quality primitives at the readout level and at high rates also creates an opportunity to perform valuable measurements in real time that were not previously possible.

\section{The VELO cluster-finding algorithm on FPGAs}
\label{sect:clustering}
\begin{figure}[H]
    \centering
    \includegraphics[width=0.7\textwidth]{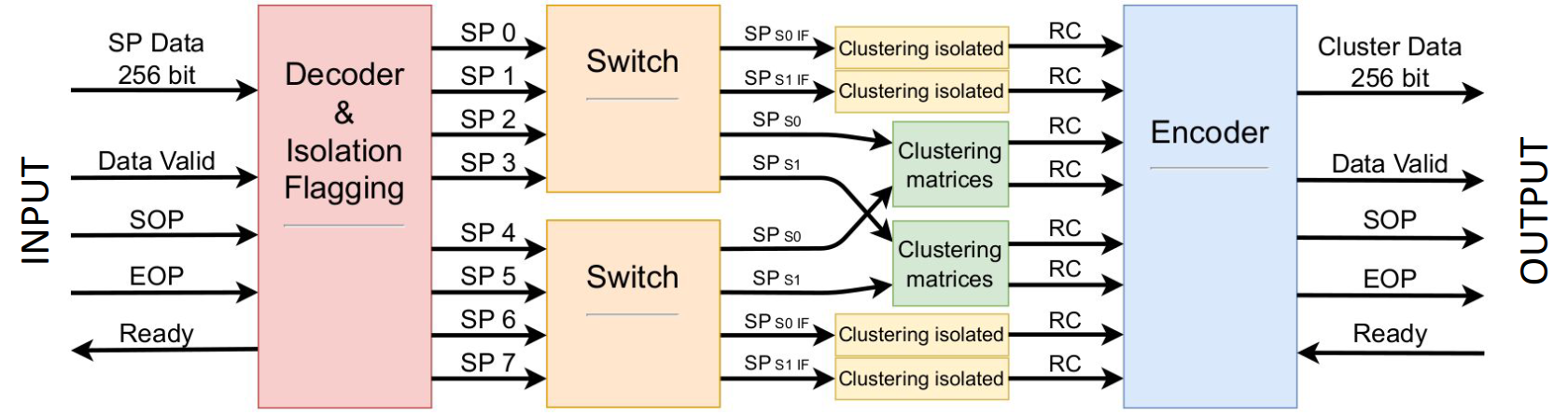}
    \caption{Key components of the VELO RetinaClustering firmware. Input SPs are decoded and flagged as isolated or not. Then two switch components arrange SPs by sensor and isolation flag. In the next stage the SPs are processed and RetinaClusters (RC) are reconstructed. At the end, all the RC are encoded and sent to the output.}
    \label{fig:clus_firmware}
\end{figure}

The LHCb Vertex Locator (VELO) is composed of 26 layers of silicon sensors, for a total of $\sim$41M digital pixels, with pitch of 55 $\mu$m.
A real-time bi-dimensional cluster-finding algorithm (RetinaClustering \cite{Bassi:2845901, 10121151}) has been developed and deployed on the Tell40 readout boards \cite{Alessio:1235873} of the VELO detector. Pixels are read out in groups of 2×4 pixels (SuperPixels, SP).
Isolated SPs are resolved very quickly with a LookUp Table. Neighboring active SPs are grouped together in a matrix-filling process and then resolved in a fully parallel way.
The clustering algorithm has been implemented in pure VHDL. The cluster-finder firmware design is composed of four main stages, concisely shown in Figure \ref{fig:clus_firmware}.

\vspace{-0.5cm}
\section{Hit counters as proxies to monitor the luminous region parameters}

A real-time monitoring of the luminosity and the luminous region parameters at the LHCb interaction point is relevant for two main reasons:
\begin{itemize}
    \item\textbf{Luminosity levelling}:
    Online luminosity estimation is needed as feedback to the LHC allowing for stable and safe experimental running conditions
    \item \textbf{Real-time alignment}:
    Knowledge of beamspot position is relevant for the experiment triggering operations
\end{itemize}
The occupancy on the VELO sensors depends on the rate of collisions \textit{i.e.} on luminosity, and on the luminous region spatial parameters. Thus, counting the number of hits provides a powerful tool to perform a real-time diagnostic of the luminous region. 

Thanks to the high granularity of the VELO detector, the occupancy on its sensors is expected to be highly linear with the instantaneous luminosity (\cite{Passaro:2842603}, Figure 5.6). Thanks to the implementation of the bidimensional clustering architecture on the readout boards of the VELO, the particle hits on the VELO are available at the very first stage of the DAQ chain.

\begin{figure}[h]
    \centering
    \subfloat[Selection Regions]{
    \includegraphics[height=0.22\textwidth]{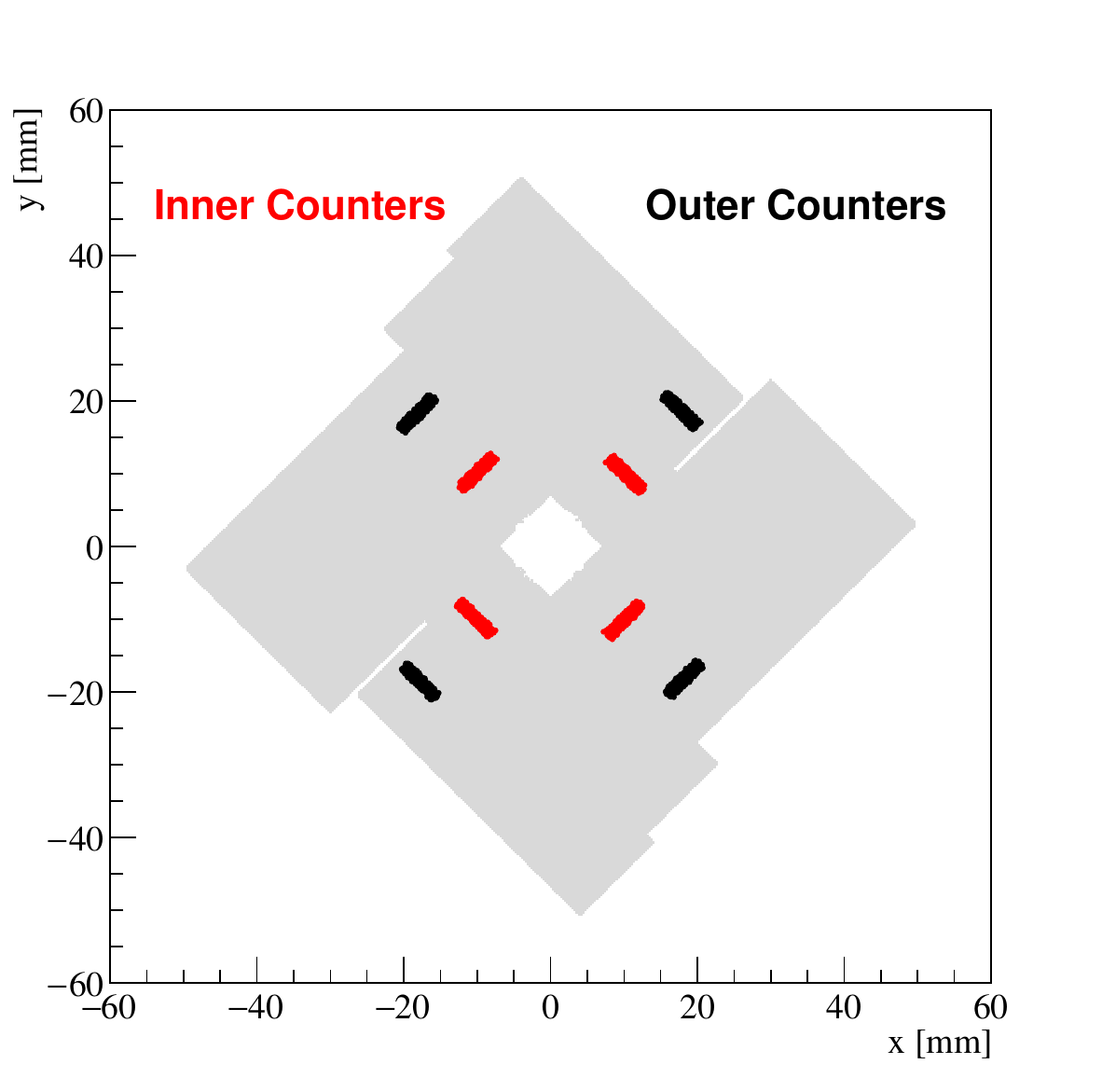}
    \label{fig:lumi_firmware_a}
    }
    \subfloat[Luminosity firmware]{
    \includegraphics[width=0.64\textwidth]{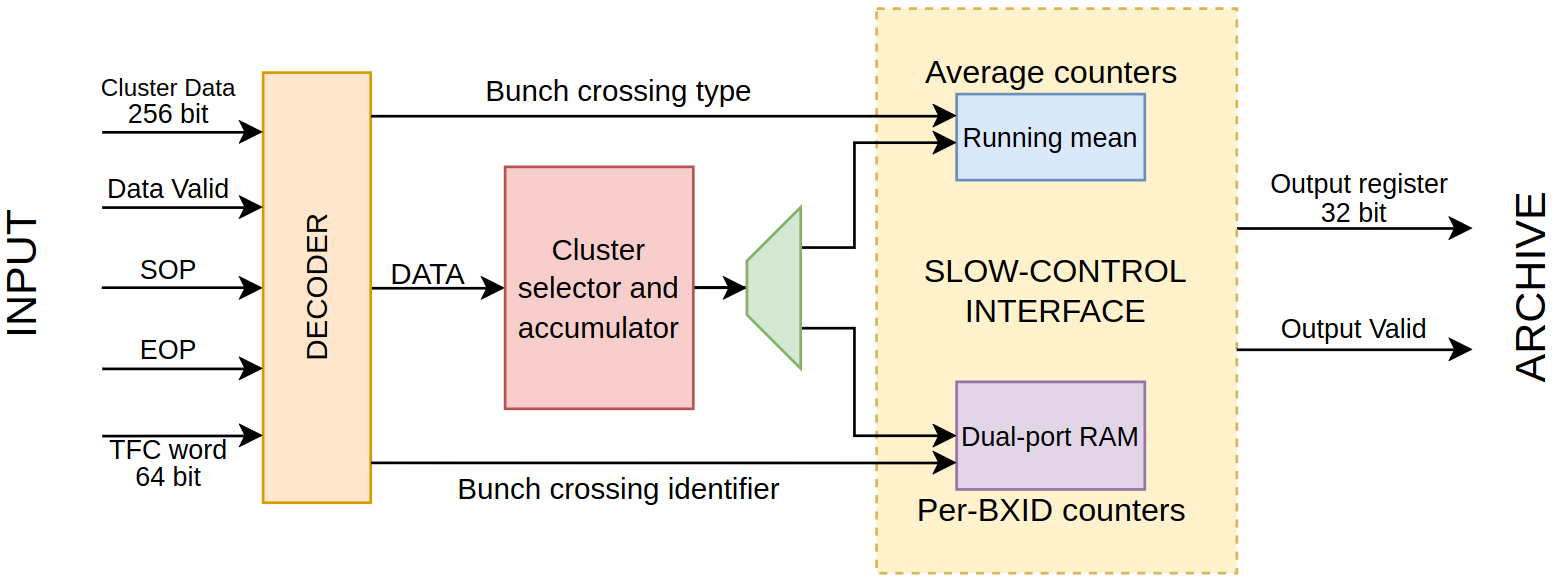}
    \label{fig:lumi_firmware}}
    \caption{Luminosity firmware implemented on the VELO FPGA readout boards. The averaged counters are computed using a running mean, which allows to cut poissonian fluctuations and to reach a better sensitivity on the luminosity estimate. The type of collision and the collision identifier are encoded in the TFC (Timing and Fast Control) word. The counters data are readout out through the Experiment Control System (ECS) and stored in an archive.}
\end{figure}

A set of \textit{luminosity counters} has been implemented on the readout boards of the VELO. On each sensor, a selection region is defined as shown in Figure \ref{fig:lumi_firmware_a}.
Linearity of such luminosity counters has been demonstrated up to average number of collisions $\mu\sim 6$ during dedicated calibration runs. For higher values of $\mu$, non-linear effects may arise, due to the FPGA-based clustering algorithm, and to the physical overlap of particle hits on the sensors. Counters divided per each colliding bunch and averaged in time have been implemented directly in the FPGA: 1) Averaged counters as feedback to the LHC beam control, readout once every 3 s; 2) Per bunch-crossing-identifier counters as feedback to the LHC injection operations, integrated over 40 s of data taking.

\noindent
Figure \ref{fig:lumi_firmware} shows the schematics of the firmware implemented to compute in real-time the hits counters. 

\section{Measuring the instantaneous luminosity}
The instantaneous luminosity is computed by measuring both the fraction of events with no hits (\textit{logZero} method) and the average occupancy (\textit{average} method) in each selection region.

\begin{figure}[h]
    \centering
    \includegraphics[width=0.65\textwidth]{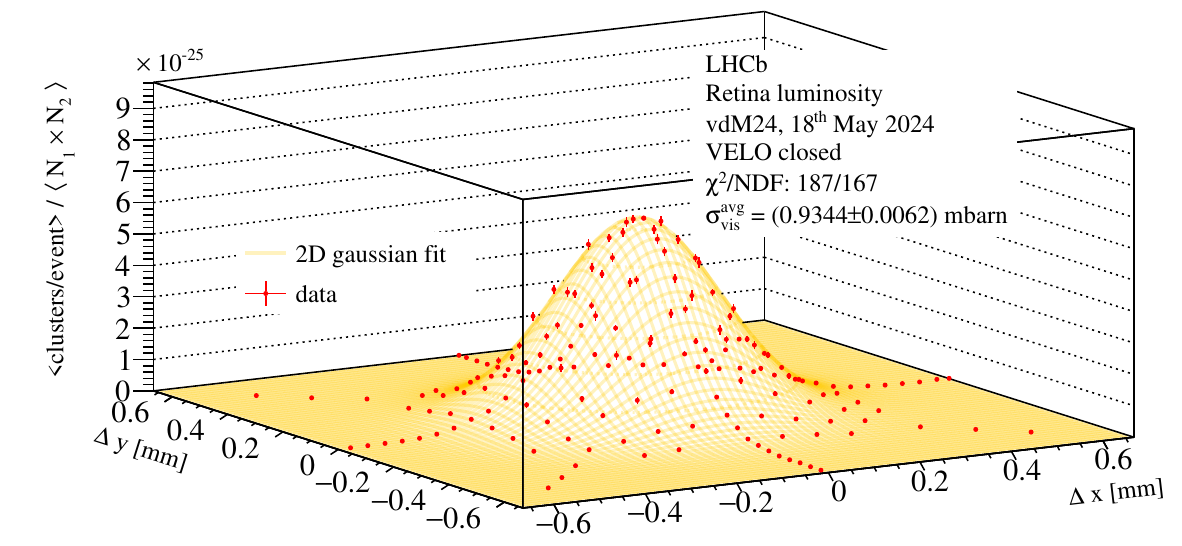}
    \caption{Example of calibration of a RetinaCluster-based luminosity counter during the May24 vdM scan \cite{LHCB-FIGURE-2024-019}. On the vertical axis the average number of hits per event, normalized by the average bunches populations product, is shown; on the horizontal axis, the beams separations on the horizontal and vertical planes are displayed. 
    The $\sigma_{\text{vis}}$ is computed as the area beneath the fitted curve.}
    \label{fig:lumi_2D_vdM_2024}
\end{figure}

Any luminosity counter needs to be calibrated in order to infer the instantaneous luminosity starting from the counter rate. This is done by measuring the \textit{visible cross-sections} $\sigma_{\text{vis}}$ via dedicated calibration runs called van der Meer (vdM) scanss \cite{Balagura_2021}. The instantaneous luminosity $\mathcal{L}_{\text{inst}}$ is related to the mean number of \textit{visible interactions} $\mu_{\text{vis}}$ by
$\mathcal{L}_{\text{inst}}= N_{\text{bb}}\dfrac{\mu_{\text{vis}}}{\sigma_{\text{vis}}}f_{\text{LHC}}$, 
where $N_{\text{bb}}$ is the number of colliding bunches and $f_{\text{LHC}}=11.245\,\text{kHz}$ is the LHC revolution frequency. The visible cross-section is a property specific to each luminosity counter, and its value and meaning depends on the method that is used to define the luminosity counter, refer to \cite{LHCB-FIGURE-2024-019} for details.
An example of calibration performed for one counter using the average method is shown in Figure \ref{fig:lumi_2D_vdM_2024}.

The calibrated luminosity counters are then averaged via a trimmed mean in order to cut out outliers and to achieve a stable and robust estimator for the instantaneous luminosity. At the nominal LHCb Run-3 luminosity ($2\cdot10^{33}\,\text{cm}^{-2}\text{s}^{.1}$), the statistical uncertainty for the trimmed luminosity estimator is <0.1\% every 3 s for the time-averaged measurement, and O(\%) for the per-bunch-crossing luminosity measurement.

\vspace{-0.5cm}
\section{Track-less luminous region position monitoring}
\begin{figure}
    \centering
    \includegraphics[width=0.5\textwidth]{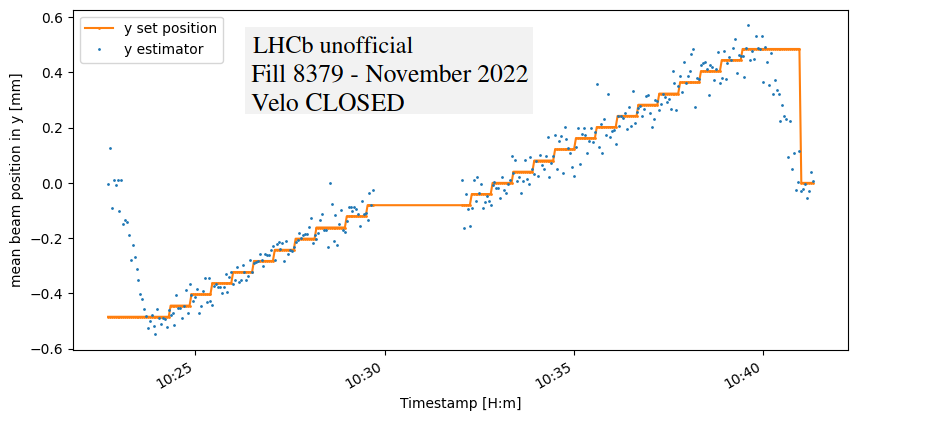}
    \caption{y-coordinate of the luminous region estimated using the PCA-based estimator. The blue points represent the y position estimated by the hit-based estimator, while the orange line is the target position as set by the LHC control system. }
    \label{fig:y_LSC_PCA}
\end{figure}

The hit distribution on the VELO is - at the first order - linear with small displacements of the luminous region position. It is possible to use the real-time hit counters to define a linear estimator for the $xyz$ mean position of the luminous region. The estimator is found by diagonalizing the covariance matrix of the hit counters. This is done by training a \textit{Principal Component Analysis} (PCA) on Monte Carlo samples with different beam positions. The components of the first eigenvector of the PCA are used as coefficients of the linear estimators. The resulting estimator is then calibrated during special LHC runs when the luminous region is displaced. An example of the luminous region position estimate computed via the PCA method is shown in Figure \ref{fig:y_LSC_PCA}.

\vspace{-.4cm}
\section{Conclusions and Future prospects}
These results represent a successful implementation of a real-time hit-statistics evaluation, transparently embedded in the readout in a complex detector at the full LHC average collision rate of 30 MHz. The real-time luminosity measurement based on the hit-counting at the readout level of the VELO is now fully integrated in the LHCb online system and it is used amongst the official LHCb luminosity tools, while the luminous region monitoring is being commissioned and integrated in the online system.
The availability of reconstructed hits at the readout level makes it possible to compute even more complex statistics:
    \bfseries 1. \mdseries Luminous region longitudinal width;
    \bfseries 2. \mdseries Luminous region inclinations in the horizontal and vertical plane;
    \bfseries 3. \mdseries Spillover (digitization of an active pixel at the next bunch crossing) and efficiency monitoring of the VELO sensors;
    \bfseries 4. \mdseries Monitoring of the relative positions of each module with respect to any other.
The LHCb collaboration is now planning to implement part of the track reconstruction using FPGAs during Run-4 (2029) (Downstream Tracker \cite{LHCbcollaboration:2886764, Lazzari:2888549}). This opens up the possibility to have a general FPGA-based approach to track reconstruction for Run-5, which will be crucial to cope with the increased instantaneous luminosity \cite{LHCbcollaboration:2886764}. The availability of reconstructed tracks at the readout level would constitute an opportunity to further improve the hit-based beam monitoring, taking full advantage of the unique real-time reconstruction capabilities of the LHCb experiment.

\section*{Acknowledgements}
These results were possible thanks to the fundings of the INFN and the kind support of the LHCb's VELO, Luminosity and Real Time Analysis (RTA) groups.

\bibliography{poster}
\bibliographystyle{unsrt}
\end{document}